\documentstyle[11pt]{article}
\def \be {\begin{equation}}
\def \ee {\end{equation}}
\def \vv{\vec{v}}

\title{A Solvable Model of a Glass}
\author{Reimer K\"uhn\thanks{supported by a Heisenberg fellowship}\\
Institut f\"ur Theoretische Physik, Universit\"at Heidelberg\\
Philosophenweg 19, D--69120 Heidelberg, Germany\\
{\small e--mail: \tt kuehn@hybrid.tphys.uni-heidelberg.de}}
\begin{document}
\maketitle
\begin{abstract}
An analytically tractable model is introduced which exhibits both, a
glass--like freezing transition, and a collection of double--well 
configurations in its zero--temperature potential energy landscape.
The latter are generally believed to be responsible for the anomalous
low--temperature properties of glass-like and amorphous systems via
a tunneling mechanism that allows particles to move back and forth
between adjacent potential energy minima. Using mean--field and replica
methods, we are able to compute the distribution of asymmetries and 
barrier--heights of the double--well configurations {\em analytically}, 
and thereby check various assumptions of the standard tunneling model. 
We find, in particular, strong correlations between asymmetries and 
barrier--heights as well as a collection of single--well configurations 
in the potential energy landscape of the glass--forming system --- 
in contrast to the assumptions of the standard model. Nevertheless, 
the specific heat scales linearly with temperature over a wide range of 
low temperatures.
\end{abstract}

\section{Introduction}

The present contribution is primarily concerned with the anomalous 
low--temperature properties of amorphous and glass-like materials. A prominent 
example of such an anomaly is the roughly linear temperature dependence of the 
specific heat at $T < 1$\,K, which is in stark contrast to the $T^3$ behaviour 
known to originate from lattice vibrations in crystaline materials. Further 
anomalies are reported for the temperature dependences of the thermal 
conductivity and other transport properties.

To explain these anomalies, a phenomenological model --- the so--called 
standard tunneling model (STM) \cite{an+,phil,jae} --- has been introduced.
It is based on two assumptions, which are plausible but until today are
still lacking an analytic foundation based on microscopic modelling. First,
it is assumed that even at temperatures well below the glass temperature,
small local rearrangements of single atoms or of small groups of atoms are
possible via tunneling between adjacent local minima in the potential energy
surface of the system. Second, individual local double--well configurations
of the potential energy surface are taken to be randomly distributed, and
a specific assumption is advanced concerning the distribution $P(\Delta,
\Delta_0)$ of asymmetries $\Delta$, and tunneling--matrix elements 
$\Delta_0$, viz., $P(\Delta,\Delta_0) \sim \Delta_0^{-1}$. The value of 
$\Delta_0$ is related with the barrier--height $V$ between adjacent minima
and the distance $d$ between them. In WKB--approximation one has
$\Delta_0 = \hbar\omega_0 \exp(-\lambda)$, with $\lambda=\frac{d}{2} \sqrt{2 
m V /\hbar^2}$. Here $\omega_0$ is a characteristic frequency (of the order of 
the frequency of harmonic oscillations in the two wells forming the double well
structure), $m$ the effective mass of the tunneling particle, and $d$ the
separation between the two minima of the double well. In terms of
$\Delta$ and $\lambda$ one has $P(\Delta,\lambda) \simeq {\rm const.}$

The STM describes experimental data reasonably well at low temperatures, i.e. for $T < 1$\,K (for an overview, see e.g. \cite{hu}). At temperatures above
1\,K, one observes a (non--phonon) $T^3$ contribution to the specific heat 
and a plateau in the thermal conductivity which cannot be accounted for within 
the set of assumptions of the STM. To model these phenomena, alternative 
assumptions concerning the distribution of local potential energy configurations have been advanced, such as those leading to the so--called soft--potential 
model \cite{kar+,bu+}, where it is assumed that locally the potential energy 
surface can be described by fourth order polynomials of the form $V(x) = u_0 
[u_2 x^2 + u_3 x^3 + x^4]$, with $u_0$ a fixed parameter and $u_2$ and $u_3$ independently distributed in a specific way. Under certain assumptions about 
these distributions, these systems also exhibit a collection of `soft'
(an)harmonic single--well potentials, supporting localized soft vibrations which can reasonably well account for both, the crossover to  $T^3$--behaviour of the 
specific heat above 1\,K as well as the plateau in the thermal conductivity.

On the other hand, simulations that tried to detect double--well potentials
in quenched Lenard-Jones mixtures \cite{heusi} produced results which did not 
fit well with the assumptions of the soft--potential model, but could be 
described by an ansatz that leads to a {\em generalized\/} soft--potential 
model, viz. $V(x) = w_2 x^2 + w_3 x^3 + w_4 x^4$, with all three coefficients 
$w_\alpha$ independently distributed in a specific way. Evaluations are, 
however, as yet based on rather moderate statistics.

For the time being, it is perhaps safe to say that both, the STM and the 
soft--potential model provide {\em phenomenological\/} descriptions, based on 
assumptions which --- while plausible in many respects --- are still lacking 
analytic support based on more microscopic approaches. 

Here we propose a microscopic model inspired by spin--glass theory which 
exhibits both, a glass--like freezing transition at a certain glass--temperature $T_{\rm g}$, and a collection of double--well configurations in its 
zero--temperature potential energy surface. Within this model, we shall not only be able to compute the full statistics of double--well configurations believed 
to be responsible for the low--temperature anomalies but also exhibit {\em 
relations\/} between low--temperature and high--temperature phenomena, e.g. 
between the low--temperature specific heat and the value of the glass--transition temperature itself.

Our line of reasoning is as follows. In Sec. 2, we propose an expression for 
the potential energy of a collection of particles forming an amorphous model 
system, which is taken to be {\em random\/} in its harmonic part, and which 
includes anharmonic on--site potentials to stabilize the system as a whole. We 
choose our setup in such a way that it can be analysed exactly within 
mean--field theory, and the replica method is used to deal with the disorder. 
By these means, the phase diagram of the model can be computed (Sec. 3), and 
we observe that it has a glass--like frozen phase at sufficiently low temperatures.

In mean--field theory, the system is described by an ensemble of effective 
single--site problems, characterized by single--site potentials which contain 
random parameters; replica theory in this context can be understood as a method 
to compute the distribution of these parameters self--consistently. The 
potential energy surface of the original model is thereby represented as 
the zero temperature limit of the aforesaid ensemble of independent 
single--site  potentials, and we are able to identify regions of parameter 
space where some members of this ensemble have double--well form. The 
distribution of the parameters characterizing the  double--well potentials
(DWPs) --- asymmetries and barrier heights --- can be computed analytically 
within replica theory (Sec. 4). Taking these distributions as input of a 
tunneling model, we are able to {\em compute\/} the contribution of the 
tunneling states to various low--temperature anomalies. In the present paper
we shall restrict our attention to the specific heat at low temperatures.  
Section 5 contains a summary  of our achievements and an outlook on open problems.
 
\section{The Model}

Consider the following expression for the potential energy of a collection 
of $N$ degrees of freedom (particles for short) forming an amorphous or
glass--like system,
\be
U_{\rm pot}(v) = -\frac{1}{2} \sum_{i,j} J_{ij} v_i v_j + \frac{1}{\gamma}
\sum_i G(v_i)\ ,
\ee
in which $v_i$ ($1\le i\le N$) may be interpreted as the deviation of the 
\mbox{$i$-th} particle from some preassigned reference position. We propose to
model the amorphous aspect of the system by taking the first, i.e. the harmonic
contribution to $U_{\rm pot}(v)$ to be {\em random\/}, so that the reference
positions and thereby the entire system would quite generally turn out to be
unstable in the harmonic approximation. This is why a set of anharmonic 
on--site potentials $G(v_i)$ is added to stabilize the system as a whole.

To be specific, we take the $J_{ij}$ to be independent Gaussians with mean $J_0/N$ and variance $J^2/N$. In order to fix the energy- and thus the temperature scale, we specialize to $J=1$ in what follows. For the on--site 
potential we choose
\be
G(v) = \frac{1}{2} v^2 + \frac{a}{4\,{\rm !}} v^4\ .
\ee
That is, $G$ also creates a harmonic restoring force, and by varying the 
parameter $\gamma$ in (1) we can tune the number of modes in the system
which are unstable at the harmonic level of description. Other forms of $G(v)$
may be contemplated; our method to solve the model does not depend on the 
particular shape of $G$. The only requirement is that it increases faster than
$v^2$ for large $v$ for the system to be stable.

The harmonic contribution to (1) is reminiscent of the SK spin--glass model
\cite{SK}, apart from the fact that we are dealing with continuous `spins' here, without any local or global constraints imposed on them. Models of the type introduced above --- albeit with different couplings and different choices for $G(v)$ --- have, however, been studied in the context of analogue neuron 
systems \cite{ho,kub}. A model with Gaussian couplings, but different $G$
has been considered by B\"os \cite{Boe}.

The choice of quenched random couplings in (1) certainly puts our model outside
the class of glass--models in the narrow sense. In view of recent ideas concerning the fundamental similarity between quenched disorder and so--called self--induced disorder as it is observed in glassy systems proper \cite{mez++},
it may nevertheless be argued that our choice should capture essential aspects
of glassy physics at low temperatures.

To analyze the potential energy surface, we compute the (configurational)
free energy of the system
\be
f_N(\beta) = -(\beta N)^{-1} \ln \int \prod_i d v_i \exp[-\beta U_{\rm 
pot}(v)]
\ee
and take its $T=0$ limit, using replica theory to average over the disorder,
so as to get {\em typical\/} results.
Standard arguments \cite{SK} give $f(\beta) = \lim_{n\to 0} f_n(\beta)$
for the quenched free energy, with
\be
n f_n(\beta) = \frac{1}{2} J_0 \sum_a m_a^2
+\frac{1}{4}\beta \sum_{a,b} q_{a,b}^2 
- \beta^{-1} \ln \int \prod_a d v^a \exp\big[-\beta U_{\rm eff}(\{v^a\})\big]\ .
\ee
Here
\be
U_{\rm eff}(\{v^a\}) = -J_0 \sum_a m_a v^a - \frac{1}{2}\beta \sum_{a,b} 
q_{ab}v^a v^b + \frac {1}{\gamma}\sum_a G(v^a)
\ee
is an effective replicated single--site potential, and the order parameters
$m_a =N^{-1} \sum_i \overline{\langle v_i^a\rangle}$ and $q_{ab} =N^{-1} \sum_i \overline{\langle v_i^a v_i^b\rangle}$ are determined as solutions of the
fixed point equations
\begin{eqnarray}
m_a & = &\langle v^a \rangle \quad ,\ a=1,\dots ,n \\
q_{ab} & = &\langle v^a v^b \rangle \quad ,\ a,b = 1,\dots ,n\ ,
\end{eqnarray}
where angular brackets denote a Gibbs average corresponding to the effective
replica potential (5), and where it is understood that the limit $n\to 0$
is eventually to be taken.

So far we have evaluated (4)--(7) only in the replica symmetric approximation
by assuming $m_a = m$ for the `polarization'-type order parameter, and $q_{aa}
= \hat q$ and  $q_{ab} = q$ for $a\ne b$ for the diagonal and off--diagonal 
entries of the matrix of Edwards-Anderson order parameters. These are determined 
from the fixed point equations
\begin{eqnarray}
m & = &\langle\, \langle v\rangle\, \rangle_z\ , \nonumber\\
\hat q & = &\langle\, \langle v^2\rangle\, \rangle_z\ , \\
q & = & \langle\, \langle v\rangle^2\, \rangle_z\ .\nonumber
\end{eqnarray}
Here $\langle\dots\rangle_z$ denotes an average over a zero-mean unit-variance
Gaussian $z$ while $\langle\dots\rangle$ without subscript is a Gibbs average 
corresponding to the effective replica-symmetric single--site potential
\be
U_{\rm RS}(v) = -[J_0 m + \sqrt{q} z] v -\frac{1}{2} C v^2 + \frac{1}{\gamma}
G(v)
\ee
with $C=\beta (\hat q - q)$.
The replica symmetric approximation thus describes a Gaussian ensemble of
independent single-site potentials $U_{\rm RS}(v)$, with para\-meters $m$, $q$
and $C$ which are determined self-consistently through (8).

\section {Phase Diagram}

The system described by (8)--(9) exhibits a glass-like freezing transition 
from an ergodic phase with $m=q=0$ to a frozen phase with $q\ne 0$ at some 
temperature $T_g$ depending on the parameters $J_0$ and $\gamma$ of the model. If $J_0$ is sufficiently large, a transition to a macroscopically polarized phase with $m\ne 0$ may also occur.
 
\begin{figure}[h]
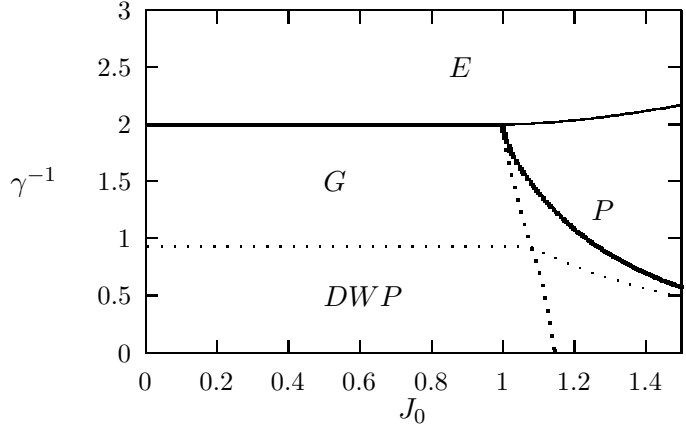

\setlength{\unitlength}{0.240900pt}
\ifx\plotpoint\undefined\newsavebox{\plotpoint}\fi
\sbox{\plotpoint}{\rule[-0.200pt]{0.400pt}{0.400pt}}%

\caption[]{$T=0$ phase diagram. $E$ denotes the $T=0$--limit of the ergodic
phase, $G$ the glassy phase, and $P$ a phase with macroscopic polarization.
The bold line is the AT line. Below the (small) dotted line is the region with
DWPs. The bigger dots separate the glassy phase $G$ from the phase $P$ with 
macroscopic polarization.}
\end{figure}

The assumption of replica symmetry is not always correct. Spontaneous replica
symmetry breaking (RSB) occurs at low temperatures (large $\beta$) and large 
$\gamma$. The precise location of the instability against RSB is given by the
AT criterion \cite{AT}
\be   
  1=\beta^2 \, \left\langle\,(\,\langle v^2\rangle - \langle v\rangle^2\,)^2\, 
\right\rangle_z =\frac{1}{q} \Big\langle \Big(\frac{{\rm d}}{{\rm d} z} \langle v\rangle
\Big)^2\, \Big \rangle_z\ ,
\ee
the second expression being more appropriate for an evaluation in the
$T=0$-limit.

Phase boundaries between ergodic and non-ergodic phases are increasing 
functions of $\gamma$, diverging as $\gamma\to\infty$, and approaching 
zero for finite values of $\gamma$ which can be read off from the $T=0$ 
phase diagram of the model in \mbox{Fig. 1}.

Interestingly, the system can also exhibit a collection of DWPs in its 
zero-temperature potential energy surface. The following Sect. is devoted to 
extracting their statistics, and to analyzing their contribution to the
low--temperature specific heat.

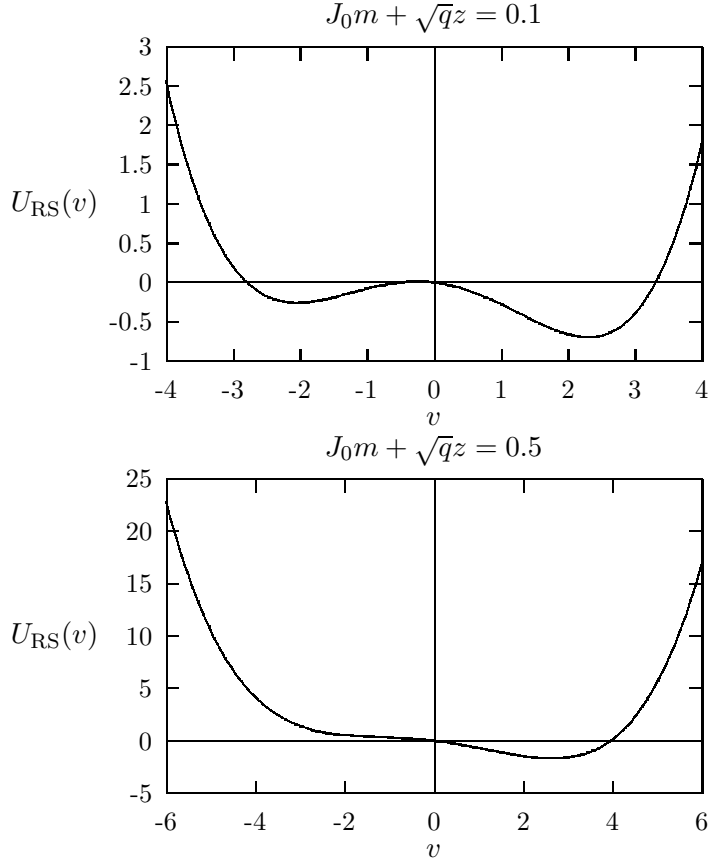
\begin{figure}[h]
\setlength{\unitlength}{0.240900pt}
\ifx\plotpoint\undefined\newsavebox{\plotpoint}\fi
\sbox{\plotpoint}{\rule[-0.200pt]{0.400pt}{0.400pt}}%
\begin{picture}(1125,675)(0,0)
\font\gnuplot=cmr10 at 10pt
\gnuplot
\sbox{\plotpoint}{\rule[-0.200pt]{0.400pt}{0.400pt}}%
\put(220.0,237.0){\rule[-0.200pt]{202.597pt}{0.400pt}}
\put(641.0,113.0){\rule[-0.200pt]{0.400pt}{119.005pt}}
\put(220.0,113.0){\rule[-0.200pt]{4.818pt}{0.400pt}}
\put(198,113){\makebox(0,0)[r]{-1}}
\put(1041.0,113.0){\rule[-0.200pt]{4.818pt}{0.400pt}}
\put(220.0,175.0){\rule[-0.200pt]{4.818pt}{0.400pt}}
\put(198,175){\makebox(0,0)[r]{-0.5}}
\put(1041.0,175.0){\rule[-0.200pt]{4.818pt}{0.400pt}}
\put(220.0,237.0){\rule[-0.200pt]{4.818pt}{0.400pt}}
\put(198,237){\makebox(0,0)[r]{0}}
\put(1041.0,237.0){\rule[-0.200pt]{4.818pt}{0.400pt}}
\put(220.0,298.0){\rule[-0.200pt]{4.818pt}{0.400pt}}
\put(198,298){\makebox(0,0)[r]{0.5}}
\put(1041.0,298.0){\rule[-0.200pt]{4.818pt}{0.400pt}}
\put(220.0,360.0){\rule[-0.200pt]{4.818pt}{0.400pt}}
\put(198,360){\makebox(0,0)[r]{1}}
\put(1041.0,360.0){\rule[-0.200pt]{4.818pt}{0.400pt}}
\put(220.0,422.0){\rule[-0.200pt]{4.818pt}{0.400pt}}
\put(198,422){\makebox(0,0)[r]{1.5}}
\put(1041.0,422.0){\rule[-0.200pt]{4.818pt}{0.400pt}}
\put(220.0,484.0){\rule[-0.200pt]{4.818pt}{0.400pt}}
\put(198,484){\makebox(0,0)[r]{2}}
\put(1041.0,484.0){\rule[-0.200pt]{4.818pt}{0.400pt}}
\put(220.0,545.0){\rule[-0.200pt]{4.818pt}{0.400pt}}
\put(198,545){\makebox(0,0)[r]{2.5}}
\put(1041.0,545.0){\rule[-0.200pt]{4.818pt}{0.400pt}}
\put(220.0,607.0){\rule[-0.200pt]{4.818pt}{0.400pt}}
\put(198,607){\makebox(0,0)[r]{3}}
\put(1041.0,607.0){\rule[-0.200pt]{4.818pt}{0.400pt}}
\put(220.0,113.0){\rule[-0.200pt]{0.400pt}{4.818pt}}
\put(220,68){\makebox(0,0){-4}}
\put(220.0,587.0){\rule[-0.200pt]{0.400pt}{4.818pt}}
\put(325.0,113.0){\rule[-0.200pt]{0.400pt}{4.818pt}}
\put(325,68){\makebox(0,0){-3}}
\put(325.0,587.0){\rule[-0.200pt]{0.400pt}{4.818pt}}
\put(430.0,113.0){\rule[-0.200pt]{0.400pt}{4.818pt}}
\put(430,68){\makebox(0,0){-2}}
\put(430.0,587.0){\rule[-0.200pt]{0.400pt}{4.818pt}}
\put(535.0,113.0){\rule[-0.200pt]{0.400pt}{4.818pt}}
\put(535,68){\makebox(0,0){-1}}
\put(535.0,587.0){\rule[-0.200pt]{0.400pt}{4.818pt}}
\put(641.0,113.0){\rule[-0.200pt]{0.400pt}{4.818pt}}
\put(641,68){\makebox(0,0){0}}
\put(641.0,587.0){\rule[-0.200pt]{0.400pt}{4.818pt}}
\put(746.0,113.0){\rule[-0.200pt]{0.400pt}{4.818pt}}
\put(746,68){\makebox(0,0){1}}
\put(746.0,587.0){\rule[-0.200pt]{0.400pt}{4.818pt}}
\put(851.0,113.0){\rule[-0.200pt]{0.400pt}{4.818pt}}
\put(851,68){\makebox(0,0){2}}
\put(851.0,587.0){\rule[-0.200pt]{0.400pt}{4.818pt}}
\put(956.0,113.0){\rule[-0.200pt]{0.400pt}{4.818pt}}
\put(956,68){\makebox(0,0){3}}
\put(956.0,587.0){\rule[-0.200pt]{0.400pt}{4.818pt}}
\put(1061.0,113.0){\rule[-0.200pt]{0.400pt}{4.818pt}}
\put(1061,68){\makebox(0,0){4}}
\put(1061.0,587.0){\rule[-0.200pt]{0.400pt}{4.818pt}}
\put(220.0,113.0){\rule[-0.200pt]{202.597pt}{0.400pt}}
\put(1061.0,113.0){\rule[-0.200pt]{0.400pt}{119.005pt}}
\put(220.0,607.0){\rule[-0.200pt]{202.597pt}{0.400pt}}
\put(45,360){\makebox(0,0){$U_{\rm RS}(v)$}}
\put(640,23){\makebox(0,0){$v$}}
\put(640,652){\makebox(0,0){$J_0 m + \sqrt{q} z =0.1$}}
\put(220.0,113.0){\rule[-0.200pt]{0.400pt}{119.005pt}}
\put(220,553){\usebox{\plotpoint}}
\multiput(220.59,544.91)(0.488,-2.409){13}{\rule{0.117pt}{1.950pt}}
\multiput(219.17,548.95)(8.000,-32.953){2}{\rule{0.400pt}{0.975pt}}
\multiput(228.59,509.31)(0.489,-1.951){15}{\rule{0.118pt}{1.611pt}}
\multiput(227.17,512.66)(9.000,-30.656){2}{\rule{0.400pt}{0.806pt}}
\multiput(237.59,474.94)(0.488,-2.079){13}{\rule{0.117pt}{1.700pt}}
\multiput(236.17,478.47)(8.000,-28.472){2}{\rule{0.400pt}{0.850pt}}
\multiput(245.59,444.23)(0.489,-1.660){15}{\rule{0.118pt}{1.389pt}}
\multiput(244.17,447.12)(9.000,-26.117){2}{\rule{0.400pt}{0.694pt}}
\multiput(254.59,415.19)(0.488,-1.682){13}{\rule{0.117pt}{1.400pt}}
\multiput(253.17,418.09)(8.000,-23.094){2}{\rule{0.400pt}{0.700pt}}
\multiput(262.59,389.97)(0.489,-1.427){15}{\rule{0.118pt}{1.211pt}}
\multiput(261.17,392.49)(9.000,-22.486){2}{\rule{0.400pt}{0.606pt}}
\multiput(271.59,365.02)(0.488,-1.418){13}{\rule{0.117pt}{1.200pt}}
\multiput(270.17,367.51)(8.000,-19.509){2}{\rule{0.400pt}{0.600pt}}
\multiput(279.59,343.90)(0.489,-1.135){15}{\rule{0.118pt}{0.989pt}}
\multiput(278.17,345.95)(9.000,-17.948){2}{\rule{0.400pt}{0.494pt}}
\multiput(288.59,323.85)(0.488,-1.154){13}{\rule{0.117pt}{1.000pt}}
\multiput(287.17,325.92)(8.000,-15.924){2}{\rule{0.400pt}{0.500pt}}
\multiput(296.59,306.63)(0.489,-0.902){15}{\rule{0.118pt}{0.811pt}}
\multiput(295.17,308.32)(9.000,-14.316){2}{\rule{0.400pt}{0.406pt}}
\multiput(305.59,290.47)(0.488,-0.956){13}{\rule{0.117pt}{0.850pt}}
\multiput(304.17,292.24)(8.000,-13.236){2}{\rule{0.400pt}{0.425pt}}
\multiput(313.59,276.19)(0.489,-0.728){15}{\rule{0.118pt}{0.678pt}}
\multiput(312.17,277.59)(9.000,-11.593){2}{\rule{0.400pt}{0.339pt}}
\multiput(322.59,263.30)(0.488,-0.692){13}{\rule{0.117pt}{0.650pt}}
\multiput(321.17,264.65)(8.000,-9.651){2}{\rule{0.400pt}{0.325pt}}
\multiput(330.59,252.74)(0.489,-0.553){15}{\rule{0.118pt}{0.544pt}}
\multiput(329.17,253.87)(9.000,-8.870){2}{\rule{0.400pt}{0.272pt}}
\multiput(339.59,242.72)(0.488,-0.560){13}{\rule{0.117pt}{0.550pt}}
\multiput(338.17,243.86)(8.000,-7.858){2}{\rule{0.400pt}{0.275pt}}
\multiput(347.00,234.93)(0.645,-0.485){11}{\rule{0.614pt}{0.117pt}}
\multiput(347.00,235.17)(7.725,-7.000){2}{\rule{0.307pt}{0.400pt}}
\multiput(356.00,227.93)(0.671,-0.482){9}{\rule{0.633pt}{0.116pt}}
\multiput(356.00,228.17)(6.685,-6.000){2}{\rule{0.317pt}{0.400pt}}
\multiput(364.00,221.93)(0.762,-0.482){9}{\rule{0.700pt}{0.116pt}}
\multiput(364.00,222.17)(7.547,-6.000){2}{\rule{0.350pt}{0.400pt}}
\multiput(373.00,215.94)(1.066,-0.468){5}{\rule{0.900pt}{0.113pt}}
\multiput(373.00,216.17)(6.132,-4.000){2}{\rule{0.450pt}{0.400pt}}
\multiput(381.00,211.95)(1.802,-0.447){3}{\rule{1.300pt}{0.108pt}}
\multiput(381.00,212.17)(6.302,-3.000){2}{\rule{0.650pt}{0.400pt}}
\put(390,208.17){\rule{1.700pt}{0.400pt}}
\multiput(390.00,209.17)(4.472,-2.000){2}{\rule{0.850pt}{0.400pt}}
\put(398,206.17){\rule{1.900pt}{0.400pt}}
\multiput(398.00,207.17)(5.056,-2.000){2}{\rule{0.950pt}{0.400pt}}
\put(407,204.67){\rule{1.927pt}{0.400pt}}
\multiput(407.00,205.17)(4.000,-1.000){2}{\rule{0.964pt}{0.400pt}}
\put(415,203.67){\rule{2.168pt}{0.400pt}}
\multiput(415.00,204.17)(4.500,-1.000){2}{\rule{1.084pt}{0.400pt}}
\put(424,203.67){\rule{1.927pt}{0.400pt}}
\multiput(424.00,203.17)(4.000,1.000){2}{\rule{0.964pt}{0.400pt}}
\put(441,204.67){\rule{1.927pt}{0.400pt}}
\multiput(441.00,204.17)(4.000,1.000){2}{\rule{0.964pt}{0.400pt}}
\put(449,206.17){\rule{1.900pt}{0.400pt}}
\multiput(449.00,205.17)(5.056,2.000){2}{\rule{0.950pt}{0.400pt}}
\put(458,207.67){\rule{1.927pt}{0.400pt}}
\multiput(458.00,207.17)(4.000,1.000){2}{\rule{0.964pt}{0.400pt}}
\put(466,209.17){\rule{1.900pt}{0.400pt}}
\multiput(466.00,208.17)(5.056,2.000){2}{\rule{0.950pt}{0.400pt}}
\put(475,211.17){\rule{1.700pt}{0.400pt}}
\multiput(475.00,210.17)(4.472,2.000){2}{\rule{0.850pt}{0.400pt}}
\put(483,213.17){\rule{1.900pt}{0.400pt}}
\multiput(483.00,212.17)(5.056,2.000){2}{\rule{0.950pt}{0.400pt}}
\multiput(492.00,215.61)(1.579,0.447){3}{\rule{1.167pt}{0.108pt}}
\multiput(492.00,214.17)(5.579,3.000){2}{\rule{0.583pt}{0.400pt}}
\put(500,218.17){\rule{1.900pt}{0.400pt}}
\multiput(500.00,217.17)(5.056,2.000){2}{\rule{0.950pt}{0.400pt}}
\put(509,220.17){\rule{1.700pt}{0.400pt}}
\multiput(509.00,219.17)(4.472,2.000){2}{\rule{0.850pt}{0.400pt}}
\put(517,222.17){\rule{1.900pt}{0.400pt}}
\multiput(517.00,221.17)(5.056,2.000){2}{\rule{0.950pt}{0.400pt}}
\multiput(526.00,224.61)(1.579,0.447){3}{\rule{1.167pt}{0.108pt}}
\multiput(526.00,223.17)(5.579,3.000){2}{\rule{0.583pt}{0.400pt}}
\put(534,227.17){\rule{1.900pt}{0.400pt}}
\multiput(534.00,226.17)(5.056,2.000){2}{\rule{0.950pt}{0.400pt}}
\put(543,229.17){\rule{1.700pt}{0.400pt}}
\multiput(543.00,228.17)(4.472,2.000){2}{\rule{0.850pt}{0.400pt}}
\put(551,230.67){\rule{2.168pt}{0.400pt}}
\multiput(551.00,230.17)(4.500,1.000){2}{\rule{1.084pt}{0.400pt}}
\put(560,232.17){\rule{1.700pt}{0.400pt}}
\multiput(560.00,231.17)(4.472,2.000){2}{\rule{0.850pt}{0.400pt}}
\put(568,233.67){\rule{2.168pt}{0.400pt}}
\multiput(568.00,233.17)(4.500,1.000){2}{\rule{1.084pt}{0.400pt}}
\put(577,234.67){\rule{1.927pt}{0.400pt}}
\multiput(577.00,234.17)(4.000,1.000){2}{\rule{0.964pt}{0.400pt}}
\put(585,235.67){\rule{2.168pt}{0.400pt}}
\multiput(585.00,235.17)(4.500,1.000){2}{\rule{1.084pt}{0.400pt}}
\put(594,236.67){\rule{1.927pt}{0.400pt}}
\multiput(594.00,236.17)(4.000,1.000){2}{\rule{0.964pt}{0.400pt}}
\put(432.0,205.0){\rule[-0.200pt]{2.168pt}{0.400pt}}
\put(628,236.67){\rule{1.927pt}{0.400pt}}
\multiput(628.00,237.17)(4.000,-1.000){2}{\rule{0.964pt}{0.400pt}}
\put(636,235.67){\rule{2.168pt}{0.400pt}}
\multiput(636.00,236.17)(4.500,-1.000){2}{\rule{1.084pt}{0.400pt}}
\put(645,234.67){\rule{1.927pt}{0.400pt}}
\multiput(645.00,235.17)(4.000,-1.000){2}{\rule{0.964pt}{0.400pt}}
\put(653,233.17){\rule{1.900pt}{0.400pt}}
\multiput(653.00,234.17)(5.056,-2.000){2}{\rule{0.950pt}{0.400pt}}
\put(662,231.17){\rule{1.700pt}{0.400pt}}
\multiput(662.00,232.17)(4.472,-2.000){2}{\rule{0.850pt}{0.400pt}}
\put(670,229.17){\rule{1.900pt}{0.400pt}}
\multiput(670.00,230.17)(5.056,-2.000){2}{\rule{0.950pt}{0.400pt}}
\multiput(679.00,227.95)(1.579,-0.447){3}{\rule{1.167pt}{0.108pt}}
\multiput(679.00,228.17)(5.579,-3.000){2}{\rule{0.583pt}{0.400pt}}
\multiput(687.00,224.95)(1.802,-0.447){3}{\rule{1.300pt}{0.108pt}}
\multiput(687.00,225.17)(6.302,-3.000){2}{\rule{0.650pt}{0.400pt}}
\multiput(696.00,221.95)(1.579,-0.447){3}{\rule{1.167pt}{0.108pt}}
\multiput(696.00,222.17)(5.579,-3.000){2}{\rule{0.583pt}{0.400pt}}
\multiput(704.00,218.95)(1.802,-0.447){3}{\rule{1.300pt}{0.108pt}}
\multiput(704.00,219.17)(6.302,-3.000){2}{\rule{0.650pt}{0.400pt}}
\multiput(713.00,215.94)(1.066,-0.468){5}{\rule{0.900pt}{0.113pt}}
\multiput(713.00,216.17)(6.132,-4.000){2}{\rule{0.450pt}{0.400pt}}
\multiput(721.00,211.95)(1.802,-0.447){3}{\rule{1.300pt}{0.108pt}}
\multiput(721.00,212.17)(6.302,-3.000){2}{\rule{0.650pt}{0.400pt}}
\multiput(730.00,208.94)(1.066,-0.468){5}{\rule{0.900pt}{0.113pt}}
\multiput(730.00,209.17)(6.132,-4.000){2}{\rule{0.450pt}{0.400pt}}
\multiput(738.00,204.94)(1.212,-0.468){5}{\rule{1.000pt}{0.113pt}}
\multiput(738.00,205.17)(6.924,-4.000){2}{\rule{0.500pt}{0.400pt}}
\multiput(747.00,200.94)(1.066,-0.468){5}{\rule{0.900pt}{0.113pt}}
\multiput(747.00,201.17)(6.132,-4.000){2}{\rule{0.450pt}{0.400pt}}
\multiput(755.00,196.93)(0.933,-0.477){7}{\rule{0.820pt}{0.115pt}}
\multiput(755.00,197.17)(7.298,-5.000){2}{\rule{0.410pt}{0.400pt}}
\multiput(764.00,191.94)(1.066,-0.468){5}{\rule{0.900pt}{0.113pt}}
\multiput(764.00,192.17)(6.132,-4.000){2}{\rule{0.450pt}{0.400pt}}
\multiput(772.00,187.94)(1.212,-0.468){5}{\rule{1.000pt}{0.113pt}}
\multiput(772.00,188.17)(6.924,-4.000){2}{\rule{0.500pt}{0.400pt}}
\multiput(781.00,183.93)(0.821,-0.477){7}{\rule{0.740pt}{0.115pt}}
\multiput(781.00,184.17)(6.464,-5.000){2}{\rule{0.370pt}{0.400pt}}
\multiput(789.00,178.94)(1.212,-0.468){5}{\rule{1.000pt}{0.113pt}}
\multiput(789.00,179.17)(6.924,-4.000){2}{\rule{0.500pt}{0.400pt}}
\multiput(798.00,174.94)(1.066,-0.468){5}{\rule{0.900pt}{0.113pt}}
\multiput(798.00,175.17)(6.132,-4.000){2}{\rule{0.450pt}{0.400pt}}
\multiput(806.00,170.94)(1.212,-0.468){5}{\rule{1.000pt}{0.113pt}}
\multiput(806.00,171.17)(6.924,-4.000){2}{\rule{0.500pt}{0.400pt}}
\multiput(815.00,166.95)(1.579,-0.447){3}{\rule{1.167pt}{0.108pt}}
\multiput(815.00,167.17)(5.579,-3.000){2}{\rule{0.583pt}{0.400pt}}
\multiput(823.00,163.94)(1.212,-0.468){5}{\rule{1.000pt}{0.113pt}}
\multiput(823.00,164.17)(6.924,-4.000){2}{\rule{0.500pt}{0.400pt}}
\multiput(832.00,159.95)(1.579,-0.447){3}{\rule{1.167pt}{0.108pt}}
\multiput(832.00,160.17)(5.579,-3.000){2}{\rule{0.583pt}{0.400pt}}
\put(840,156.17){\rule{1.900pt}{0.400pt}}
\multiput(840.00,157.17)(5.056,-2.000){2}{\rule{0.950pt}{0.400pt}}
\put(849,154.17){\rule{1.700pt}{0.400pt}}
\multiput(849.00,155.17)(4.472,-2.000){2}{\rule{0.850pt}{0.400pt}}
\put(857,152.17){\rule{1.900pt}{0.400pt}}
\multiput(857.00,153.17)(5.056,-2.000){2}{\rule{0.950pt}{0.400pt}}
\put(866,150.67){\rule{1.927pt}{0.400pt}}
\multiput(866.00,151.17)(4.000,-1.000){2}{\rule{0.964pt}{0.400pt}}
\put(602.0,238.0){\rule[-0.200pt]{6.263pt}{0.400pt}}
\put(891,150.67){\rule{2.168pt}{0.400pt}}
\multiput(891.00,150.17)(4.500,1.000){2}{\rule{1.084pt}{0.400pt}}
\put(900,152.17){\rule{1.700pt}{0.400pt}}
\multiput(900.00,151.17)(4.472,2.000){2}{\rule{0.850pt}{0.400pt}}
\multiput(908.00,154.60)(1.212,0.468){5}{\rule{1.000pt}{0.113pt}}
\multiput(908.00,153.17)(6.924,4.000){2}{\rule{0.500pt}{0.400pt}}
\multiput(917.00,158.60)(1.066,0.468){5}{\rule{0.900pt}{0.113pt}}
\multiput(917.00,157.17)(6.132,4.000){2}{\rule{0.450pt}{0.400pt}}
\multiput(925.00,162.59)(0.933,0.477){7}{\rule{0.820pt}{0.115pt}}
\multiput(925.00,161.17)(7.298,5.000){2}{\rule{0.410pt}{0.400pt}}
\multiput(934.00,167.59)(0.569,0.485){11}{\rule{0.557pt}{0.117pt}}
\multiput(934.00,166.17)(6.844,7.000){2}{\rule{0.279pt}{0.400pt}}
\multiput(942.00,174.59)(0.560,0.488){13}{\rule{0.550pt}{0.117pt}}
\multiput(942.00,173.17)(7.858,8.000){2}{\rule{0.275pt}{0.400pt}}
\multiput(951.59,182.00)(0.488,0.560){13}{\rule{0.117pt}{0.550pt}}
\multiput(950.17,182.00)(8.000,7.858){2}{\rule{0.400pt}{0.275pt}}
\multiput(959.59,191.00)(0.489,0.611){15}{\rule{0.118pt}{0.589pt}}
\multiput(958.17,191.00)(9.000,9.778){2}{\rule{0.400pt}{0.294pt}}
\multiput(968.59,202.00)(0.488,0.824){13}{\rule{0.117pt}{0.750pt}}
\multiput(967.17,202.00)(8.000,11.443){2}{\rule{0.400pt}{0.375pt}}
\multiput(976.59,215.00)(0.489,0.786){15}{\rule{0.118pt}{0.722pt}}
\multiput(975.17,215.00)(9.000,12.501){2}{\rule{0.400pt}{0.361pt}}
\multiput(985.59,229.00)(0.488,1.022){13}{\rule{0.117pt}{0.900pt}}
\multiput(984.17,229.00)(8.000,14.132){2}{\rule{0.400pt}{0.450pt}}
\multiput(993.59,245.00)(0.489,1.019){15}{\rule{0.118pt}{0.900pt}}
\multiput(992.17,245.00)(9.000,16.132){2}{\rule{0.400pt}{0.450pt}}
\multiput(1002.59,263.00)(0.488,1.286){13}{\rule{0.117pt}{1.100pt}}
\multiput(1001.17,263.00)(8.000,17.717){2}{\rule{0.400pt}{0.550pt}}
\multiput(1010.59,283.00)(0.489,1.310){15}{\rule{0.118pt}{1.122pt}}
\multiput(1009.17,283.00)(9.000,20.671){2}{\rule{0.400pt}{0.561pt}}
\multiput(1019.59,306.00)(0.488,1.550){13}{\rule{0.117pt}{1.300pt}}
\multiput(1018.17,306.00)(8.000,21.302){2}{\rule{0.400pt}{0.650pt}}
\multiput(1027.59,330.00)(0.489,1.543){15}{\rule{0.118pt}{1.300pt}}
\multiput(1026.17,330.00)(9.000,24.302){2}{\rule{0.400pt}{0.650pt}}
\multiput(1036.59,357.00)(0.488,1.947){13}{\rule{0.117pt}{1.600pt}}
\multiput(1035.17,357.00)(8.000,26.679){2}{\rule{0.400pt}{0.800pt}}
\multiput(1044.59,387.00)(0.489,1.834){15}{\rule{0.118pt}{1.522pt}}
\multiput(1043.17,387.00)(9.000,28.841){2}{\rule{0.400pt}{0.761pt}}
\multiput(1053.59,419.00)(0.488,2.277){13}{\rule{0.117pt}{1.850pt}}
\multiput(1052.17,419.00)(8.000,31.160){2}{\rule{0.400pt}{0.925pt}}
\put(874.0,151.0){\rule[-0.200pt]{4.095pt}{0.400pt}}
\end{picture}

\setlength{\unitlength}{0.240900pt}
\ifx\plotpoint\undefined\newsavebox{\plotpoint}\fi
\sbox{\plotpoint}{\rule[-0.200pt]{0.400pt}{0.400pt}}%
\begin{picture}(1125,675)(0,0)
\font\gnuplot=cmr10 at 10pt
\gnuplot
\sbox{\plotpoint}{\rule[-0.200pt]{0.400pt}{0.400pt}}%
\put(220.0,195.0){\rule[-0.200pt]{202.597pt}{0.400pt}}
\put(641.0,113.0){\rule[-0.200pt]{0.400pt}{119.005pt}}
\put(220.0,113.0){\rule[-0.200pt]{4.818pt}{0.400pt}}
\put(198,113){\makebox(0,0)[r]{-5}}
\put(1041.0,113.0){\rule[-0.200pt]{4.818pt}{0.400pt}}
\put(220.0,195.0){\rule[-0.200pt]{4.818pt}{0.400pt}}
\put(198,195){\makebox(0,0)[r]{0}}
\put(1041.0,195.0){\rule[-0.200pt]{4.818pt}{0.400pt}}
\put(220.0,278.0){\rule[-0.200pt]{4.818pt}{0.400pt}}
\put(198,278){\makebox(0,0)[r]{5}}
\put(1041.0,278.0){\rule[-0.200pt]{4.818pt}{0.400pt}}
\put(220.0,360.0){\rule[-0.200pt]{4.818pt}{0.400pt}}
\put(198,360){\makebox(0,0)[r]{10}}
\put(1041.0,360.0){\rule[-0.200pt]{4.818pt}{0.400pt}}
\put(220.0,442.0){\rule[-0.200pt]{4.818pt}{0.400pt}}
\put(198,442){\makebox(0,0)[r]{15}}
\put(1041.0,442.0){\rule[-0.200pt]{4.818pt}{0.400pt}}
\put(220.0,525.0){\rule[-0.200pt]{4.818pt}{0.400pt}}
\put(198,525){\makebox(0,0)[r]{20}}
\put(1041.0,525.0){\rule[-0.200pt]{4.818pt}{0.400pt}}
\put(220.0,607.0){\rule[-0.200pt]{4.818pt}{0.400pt}}
\put(198,607){\makebox(0,0)[r]{25}}
\put(1041.0,607.0){\rule[-0.200pt]{4.818pt}{0.400pt}}
\put(220.0,113.0){\rule[-0.200pt]{0.400pt}{4.818pt}}
\put(220,68){\makebox(0,0){-6}}
\put(220.0,587.0){\rule[-0.200pt]{0.400pt}{4.818pt}}
\put(360.0,113.0){\rule[-0.200pt]{0.400pt}{4.818pt}}
\put(360,68){\makebox(0,0){-4}}
\put(360.0,587.0){\rule[-0.200pt]{0.400pt}{4.818pt}}
\put(500.0,113.0){\rule[-0.200pt]{0.400pt}{4.818pt}}
\put(500,68){\makebox(0,0){-2}}
\put(500.0,587.0){\rule[-0.200pt]{0.400pt}{4.818pt}}
\put(641.0,113.0){\rule[-0.200pt]{0.400pt}{4.818pt}}
\put(641,68){\makebox(0,0){0}}
\put(641.0,587.0){\rule[-0.200pt]{0.400pt}{4.818pt}}
\put(781.0,113.0){\rule[-0.200pt]{0.400pt}{4.818pt}}
\put(781,68){\makebox(0,0){2}}
\put(781.0,587.0){\rule[-0.200pt]{0.400pt}{4.818pt}}
\put(921.0,113.0){\rule[-0.200pt]{0.400pt}{4.818pt}}
\put(921,68){\makebox(0,0){4}}
\put(921.0,587.0){\rule[-0.200pt]{0.400pt}{4.818pt}}
\put(1061.0,113.0){\rule[-0.200pt]{0.400pt}{4.818pt}}
\put(1061,68){\makebox(0,0){6}}
\put(1061.0,587.0){\rule[-0.200pt]{0.400pt}{4.818pt}}
\put(220.0,113.0){\rule[-0.200pt]{202.597pt}{0.400pt}}
\put(1061.0,113.0){\rule[-0.200pt]{0.400pt}{119.005pt}}
\put(220.0,607.0){\rule[-0.200pt]{202.597pt}{0.400pt}}
\put(45,360){\makebox(0,0){$U_{\rm RS}(v)$}}
\put(640,23){\makebox(0,0){$v$}}
\put(640,652){\makebox(0,0){$J_0 m + \sqrt{q} z =0.5$}}
\put(220.0,113.0){\rule[-0.200pt]{0.400pt}{119.005pt}}
\put(220,572){\usebox{\plotpoint}}
\multiput(220.59,565.15)(0.488,-2.013){13}{\rule{0.117pt}{1.650pt}}
\multiput(219.17,568.58)(8.000,-27.575){2}{\rule{0.400pt}{0.825pt}}
\multiput(228.59,535.23)(0.489,-1.660){15}{\rule{0.118pt}{1.389pt}}
\multiput(227.17,538.12)(9.000,-26.117){2}{\rule{0.400pt}{0.694pt}}
\multiput(237.59,505.77)(0.488,-1.814){13}{\rule{0.117pt}{1.500pt}}
\multiput(236.17,508.89)(8.000,-24.887){2}{\rule{0.400pt}{0.750pt}}
\multiput(245.59,478.97)(0.489,-1.427){15}{\rule{0.118pt}{1.211pt}}
\multiput(244.17,481.49)(9.000,-22.486){2}{\rule{0.400pt}{0.606pt}}
\multiput(254.59,453.60)(0.488,-1.550){13}{\rule{0.117pt}{1.300pt}}
\multiput(253.17,456.30)(8.000,-21.302){2}{\rule{0.400pt}{0.650pt}}
\multiput(262.59,430.53)(0.489,-1.252){15}{\rule{0.118pt}{1.078pt}}
\multiput(261.17,432.76)(9.000,-19.763){2}{\rule{0.400pt}{0.539pt}}
\multiput(271.59,408.43)(0.488,-1.286){13}{\rule{0.117pt}{1.100pt}}
\multiput(270.17,410.72)(8.000,-17.717){2}{\rule{0.400pt}{0.550pt}}
\multiput(279.59,389.08)(0.489,-1.077){15}{\rule{0.118pt}{0.944pt}}
\multiput(278.17,391.04)(9.000,-17.040){2}{\rule{0.400pt}{0.472pt}}
\multiput(288.59,369.85)(0.488,-1.154){13}{\rule{0.117pt}{1.000pt}}
\multiput(287.17,371.92)(8.000,-15.924){2}{\rule{0.400pt}{0.500pt}}
\multiput(296.59,352.63)(0.489,-0.902){15}{\rule{0.118pt}{0.811pt}}
\multiput(295.17,354.32)(9.000,-14.316){2}{\rule{0.400pt}{0.406pt}}
\multiput(305.59,336.47)(0.488,-0.956){13}{\rule{0.117pt}{0.850pt}}
\multiput(304.17,338.24)(8.000,-13.236){2}{\rule{0.400pt}{0.425pt}}
\multiput(313.59,322.19)(0.489,-0.728){15}{\rule{0.118pt}{0.678pt}}
\multiput(312.17,323.59)(9.000,-11.593){2}{\rule{0.400pt}{0.339pt}}
\multiput(322.59,308.89)(0.488,-0.824){13}{\rule{0.117pt}{0.750pt}}
\multiput(321.17,310.44)(8.000,-11.443){2}{\rule{0.400pt}{0.375pt}}
\multiput(330.59,296.56)(0.489,-0.611){15}{\rule{0.118pt}{0.589pt}}
\multiput(329.17,297.78)(9.000,-9.778){2}{\rule{0.400pt}{0.294pt}}
\multiput(339.59,285.51)(0.488,-0.626){13}{\rule{0.117pt}{0.600pt}}
\multiput(338.17,286.75)(8.000,-8.755){2}{\rule{0.400pt}{0.300pt}}
\multiput(347.59,275.74)(0.489,-0.553){15}{\rule{0.118pt}{0.544pt}}
\multiput(346.17,276.87)(9.000,-8.870){2}{\rule{0.400pt}{0.272pt}}
\multiput(356.00,266.93)(0.494,-0.488){13}{\rule{0.500pt}{0.117pt}}
\multiput(356.00,267.17)(6.962,-8.000){2}{\rule{0.250pt}{0.400pt}}
\multiput(364.00,258.93)(0.560,-0.488){13}{\rule{0.550pt}{0.117pt}}
\multiput(364.00,259.17)(7.858,-8.000){2}{\rule{0.275pt}{0.400pt}}
\multiput(373.00,250.93)(0.569,-0.485){11}{\rule{0.557pt}{0.117pt}}
\multiput(373.00,251.17)(6.844,-7.000){2}{\rule{0.279pt}{0.400pt}}
\multiput(381.00,243.93)(0.762,-0.482){9}{\rule{0.700pt}{0.116pt}}
\multiput(381.00,244.17)(7.547,-6.000){2}{\rule{0.350pt}{0.400pt}}
\multiput(390.00,237.93)(0.821,-0.477){7}{\rule{0.740pt}{0.115pt}}
\multiput(390.00,238.17)(6.464,-5.000){2}{\rule{0.370pt}{0.400pt}}
\multiput(398.00,232.93)(0.933,-0.477){7}{\rule{0.820pt}{0.115pt}}
\multiput(398.00,233.17)(7.298,-5.000){2}{\rule{0.410pt}{0.400pt}}
\multiput(407.00,227.94)(1.066,-0.468){5}{\rule{0.900pt}{0.113pt}}
\multiput(407.00,228.17)(6.132,-4.000){2}{\rule{0.450pt}{0.400pt}}
\multiput(415.00,223.94)(1.212,-0.468){5}{\rule{1.000pt}{0.113pt}}
\multiput(415.00,224.17)(6.924,-4.000){2}{\rule{0.500pt}{0.400pt}}
\multiput(424.00,219.95)(1.579,-0.447){3}{\rule{1.167pt}{0.108pt}}
\multiput(424.00,220.17)(5.579,-3.000){2}{\rule{0.583pt}{0.400pt}}
\multiput(432.00,216.95)(1.802,-0.447){3}{\rule{1.300pt}{0.108pt}}
\multiput(432.00,217.17)(6.302,-3.000){2}{\rule{0.650pt}{0.400pt}}
\multiput(441.00,213.95)(1.579,-0.447){3}{\rule{1.167pt}{0.108pt}}
\multiput(441.00,214.17)(5.579,-3.000){2}{\rule{0.583pt}{0.400pt}}
\put(449,210.17){\rule{1.900pt}{0.400pt}}
\multiput(449.00,211.17)(5.056,-2.000){2}{\rule{0.950pt}{0.400pt}}
\put(458,208.67){\rule{1.927pt}{0.400pt}}
\multiput(458.00,209.17)(4.000,-1.000){2}{\rule{0.964pt}{0.400pt}}
\put(466,207.17){\rule{1.900pt}{0.400pt}}
\multiput(466.00,208.17)(5.056,-2.000){2}{\rule{0.950pt}{0.400pt}}
\put(475,205.67){\rule{1.927pt}{0.400pt}}
\multiput(475.00,206.17)(4.000,-1.000){2}{\rule{0.964pt}{0.400pt}}
\put(483,204.67){\rule{2.168pt}{0.400pt}}
\multiput(483.00,205.17)(4.500,-1.000){2}{\rule{1.084pt}{0.400pt}}
\put(492,203.67){\rule{1.927pt}{0.400pt}}
\multiput(492.00,204.17)(4.000,-1.000){2}{\rule{0.964pt}{0.400pt}}
\put(509,202.67){\rule{1.927pt}{0.400pt}}
\multiput(509.00,203.17)(4.000,-1.000){2}{\rule{0.964pt}{0.400pt}}
\put(500.0,204.0){\rule[-0.200pt]{2.168pt}{0.400pt}}
\put(526,201.67){\rule{1.927pt}{0.400pt}}
\multiput(526.00,202.17)(4.000,-1.000){2}{\rule{0.964pt}{0.400pt}}
\put(517.0,203.0){\rule[-0.200pt]{2.168pt}{0.400pt}}
\put(543,200.67){\rule{1.927pt}{0.400pt}}
\multiput(543.00,201.17)(4.000,-1.000){2}{\rule{0.964pt}{0.400pt}}
\put(534.0,202.0){\rule[-0.200pt]{2.168pt}{0.400pt}}
\put(568,199.67){\rule{2.168pt}{0.400pt}}
\multiput(568.00,200.17)(4.500,-1.000){2}{\rule{1.084pt}{0.400pt}}
\put(551.0,201.0){\rule[-0.200pt]{4.095pt}{0.400pt}}
\put(585,198.67){\rule{2.168pt}{0.400pt}}
\multiput(585.00,199.17)(4.500,-1.000){2}{\rule{1.084pt}{0.400pt}}
\put(577.0,200.0){\rule[-0.200pt]{1.927pt}{0.400pt}}
\put(602,197.67){\rule{2.168pt}{0.400pt}}
\multiput(602.00,198.17)(4.500,-1.000){2}{\rule{1.084pt}{0.400pt}}
\put(594.0,199.0){\rule[-0.200pt]{1.927pt}{0.400pt}}
\put(619,196.67){\rule{2.168pt}{0.400pt}}
\multiput(619.00,197.17)(4.500,-1.000){2}{\rule{1.084pt}{0.400pt}}
\put(628,195.67){\rule{1.927pt}{0.400pt}}
\multiput(628.00,196.17)(4.000,-1.000){2}{\rule{0.964pt}{0.400pt}}
\put(636,194.67){\rule{2.168pt}{0.400pt}}
\multiput(636.00,195.17)(4.500,-1.000){2}{\rule{1.084pt}{0.400pt}}
\put(645,193.67){\rule{1.927pt}{0.400pt}}
\multiput(645.00,194.17)(4.000,-1.000){2}{\rule{0.964pt}{0.400pt}}
\put(653,192.67){\rule{2.168pt}{0.400pt}}
\multiput(653.00,193.17)(4.500,-1.000){2}{\rule{1.084pt}{0.400pt}}
\put(662,191.17){\rule{1.700pt}{0.400pt}}
\multiput(662.00,192.17)(4.472,-2.000){2}{\rule{0.850pt}{0.400pt}}
\put(670,189.67){\rule{2.168pt}{0.400pt}}
\multiput(670.00,190.17)(4.500,-1.000){2}{\rule{1.084pt}{0.400pt}}
\put(679,188.17){\rule{1.700pt}{0.400pt}}
\multiput(679.00,189.17)(4.472,-2.000){2}{\rule{0.850pt}{0.400pt}}
\put(687,186.67){\rule{2.168pt}{0.400pt}}
\multiput(687.00,187.17)(4.500,-1.000){2}{\rule{1.084pt}{0.400pt}}
\put(696,185.17){\rule{1.700pt}{0.400pt}}
\multiput(696.00,186.17)(4.472,-2.000){2}{\rule{0.850pt}{0.400pt}}
\put(704,183.67){\rule{2.168pt}{0.400pt}}
\multiput(704.00,184.17)(4.500,-1.000){2}{\rule{1.084pt}{0.400pt}}
\put(713,182.17){\rule{1.700pt}{0.400pt}}
\multiput(713.00,183.17)(4.472,-2.000){2}{\rule{0.850pt}{0.400pt}}
\put(721,180.17){\rule{1.900pt}{0.400pt}}
\multiput(721.00,181.17)(5.056,-2.000){2}{\rule{0.950pt}{0.400pt}}
\put(730,178.67){\rule{1.927pt}{0.400pt}}
\multiput(730.00,179.17)(4.000,-1.000){2}{\rule{0.964pt}{0.400pt}}
\put(738,177.17){\rule{1.900pt}{0.400pt}}
\multiput(738.00,178.17)(5.056,-2.000){2}{\rule{0.950pt}{0.400pt}}
\put(747,175.67){\rule{1.927pt}{0.400pt}}
\multiput(747.00,176.17)(4.000,-1.000){2}{\rule{0.964pt}{0.400pt}}
\put(755,174.17){\rule{1.900pt}{0.400pt}}
\multiput(755.00,175.17)(5.056,-2.000){2}{\rule{0.950pt}{0.400pt}}
\put(764,172.67){\rule{1.927pt}{0.400pt}}
\multiput(764.00,173.17)(4.000,-1.000){2}{\rule{0.964pt}{0.400pt}}
\put(772,171.17){\rule{1.900pt}{0.400pt}}
\multiput(772.00,172.17)(5.056,-2.000){2}{\rule{0.950pt}{0.400pt}}
\put(781,169.67){\rule{1.927pt}{0.400pt}}
\multiput(781.00,170.17)(4.000,-1.000){2}{\rule{0.964pt}{0.400pt}}
\put(789,168.67){\rule{2.168pt}{0.400pt}}
\multiput(789.00,169.17)(4.500,-1.000){2}{\rule{1.084pt}{0.400pt}}
\put(798,167.67){\rule{1.927pt}{0.400pt}}
\multiput(798.00,168.17)(4.000,-1.000){2}{\rule{0.964pt}{0.400pt}}
\put(611.0,198.0){\rule[-0.200pt]{1.927pt}{0.400pt}}
\put(840,167.67){\rule{2.168pt}{0.400pt}}
\multiput(840.00,167.17)(4.500,1.000){2}{\rule{1.084pt}{0.400pt}}
\put(849,168.67){\rule{1.927pt}{0.400pt}}
\multiput(849.00,168.17)(4.000,1.000){2}{\rule{0.964pt}{0.400pt}}
\put(857,170.17){\rule{1.900pt}{0.400pt}}
\multiput(857.00,169.17)(5.056,2.000){2}{\rule{0.950pt}{0.400pt}}
\put(866,172.17){\rule{1.700pt}{0.400pt}}
\multiput(866.00,171.17)(4.472,2.000){2}{\rule{0.850pt}{0.400pt}}
\multiput(874.00,174.61)(1.802,0.447){3}{\rule{1.300pt}{0.108pt}}
\multiput(874.00,173.17)(6.302,3.000){2}{\rule{0.650pt}{0.400pt}}
\multiput(883.00,177.61)(1.579,0.447){3}{\rule{1.167pt}{0.108pt}}
\multiput(883.00,176.17)(5.579,3.000){2}{\rule{0.583pt}{0.400pt}}
\multiput(891.00,180.60)(1.212,0.468){5}{\rule{1.000pt}{0.113pt}}
\multiput(891.00,179.17)(6.924,4.000){2}{\rule{0.500pt}{0.400pt}}
\multiput(900.00,184.59)(0.821,0.477){7}{\rule{0.740pt}{0.115pt}}
\multiput(900.00,183.17)(6.464,5.000){2}{\rule{0.370pt}{0.400pt}}
\multiput(908.00,189.59)(0.762,0.482){9}{\rule{0.700pt}{0.116pt}}
\multiput(908.00,188.17)(7.547,6.000){2}{\rule{0.350pt}{0.400pt}}
\multiput(917.00,195.59)(0.671,0.482){9}{\rule{0.633pt}{0.116pt}}
\multiput(917.00,194.17)(6.685,6.000){2}{\rule{0.317pt}{0.400pt}}
\multiput(925.00,201.59)(0.560,0.488){13}{\rule{0.550pt}{0.117pt}}
\multiput(925.00,200.17)(7.858,8.000){2}{\rule{0.275pt}{0.400pt}}
\multiput(934.00,209.59)(0.494,0.488){13}{\rule{0.500pt}{0.117pt}}
\multiput(934.00,208.17)(6.962,8.000){2}{\rule{0.250pt}{0.400pt}}
\multiput(942.00,217.59)(0.495,0.489){15}{\rule{0.500pt}{0.118pt}}
\multiput(942.00,216.17)(7.962,9.000){2}{\rule{0.250pt}{0.400pt}}
\multiput(951.59,226.00)(0.488,0.692){13}{\rule{0.117pt}{0.650pt}}
\multiput(950.17,226.00)(8.000,9.651){2}{\rule{0.400pt}{0.325pt}}
\multiput(959.59,237.00)(0.489,0.669){15}{\rule{0.118pt}{0.633pt}}
\multiput(958.17,237.00)(9.000,10.685){2}{\rule{0.400pt}{0.317pt}}
\multiput(968.59,249.00)(0.488,0.758){13}{\rule{0.117pt}{0.700pt}}
\multiput(967.17,249.00)(8.000,10.547){2}{\rule{0.400pt}{0.350pt}}
\multiput(976.59,261.00)(0.489,0.844){15}{\rule{0.118pt}{0.767pt}}
\multiput(975.17,261.00)(9.000,13.409){2}{\rule{0.400pt}{0.383pt}}
\multiput(985.59,276.00)(0.488,0.956){13}{\rule{0.117pt}{0.850pt}}
\multiput(984.17,276.00)(8.000,13.236){2}{\rule{0.400pt}{0.425pt}}
\multiput(993.59,291.00)(0.489,0.961){15}{\rule{0.118pt}{0.856pt}}
\multiput(992.17,291.00)(9.000,15.224){2}{\rule{0.400pt}{0.428pt}}
\multiput(1002.59,308.00)(0.488,1.154){13}{\rule{0.117pt}{1.000pt}}
\multiput(1001.17,308.00)(8.000,15.924){2}{\rule{0.400pt}{0.500pt}}
\multiput(1010.59,326.00)(0.489,1.135){15}{\rule{0.118pt}{0.989pt}}
\multiput(1009.17,326.00)(9.000,17.948){2}{\rule{0.400pt}{0.494pt}}
\multiput(1019.59,346.00)(0.488,1.418){13}{\rule{0.117pt}{1.200pt}}
\multiput(1018.17,346.00)(8.000,19.509){2}{\rule{0.400pt}{0.600pt}}
\multiput(1027.59,368.00)(0.489,1.310){15}{\rule{0.118pt}{1.122pt}}
\multiput(1026.17,368.00)(9.000,20.671){2}{\rule{0.400pt}{0.561pt}}
\multiput(1036.59,391.00)(0.488,1.682){13}{\rule{0.117pt}{1.400pt}}
\multiput(1035.17,391.00)(8.000,23.094){2}{\rule{0.400pt}{0.700pt}}
\multiput(1044.59,417.00)(0.489,1.543){15}{\rule{0.118pt}{1.300pt}}
\multiput(1043.17,417.00)(9.000,24.302){2}{\rule{0.400pt}{0.650pt}}
\multiput(1053.59,444.00)(0.488,1.880){13}{\rule{0.117pt}{1.550pt}}
\multiput(1052.17,444.00)(8.000,25.783){2}{\rule{0.400pt}{0.775pt}}
\put(806.0,168.0){\rule[-0.200pt]{8.191pt}{0.400pt}}
\end{picture}
\caption[]{ Effective replica--symmetric single--site potential for two 
different values of $\tilde h = J_0 m +\sqrt q z$. {\bf (a)} $\tilde h =0.1$.
{\bf (b)} $\tilde h =0.5$.  In {\bf (a)}, the asymmetry $\Delta$ is the difference between the two minima, the barrier height $V$ is the difference
between the maximum and the (lower) minimum, and $d$ is the separation between
minima on the $v$ axis. In both cases we have $\gamma^{-1}=0.5$ and $J_0$ such 
that $m = 0$.}
\end{figure}

\section{Double--Well Potentials and  Specific Heat}

We take the $T\to 0$  ($\beta\to\infty$) limit of the above set (8) of 
fixed point equations to compute the $T\to 0$ limit of the order parameters characterizing the ensemble (9) of effective replica symmetric single--site potentials, which in turn represents the potential energy surface of our
model (in the RS approximation). For suitable values of external parameters, some of these potentials have DWP form. From (2),(9) it is clear that the 
condition for this to occur is $\gamma C > 1$; this region is marked DWP
in Fig. 1. For not too large $|z|$ then, DWPs occur and the distribution of 
their characteristic parameters (barrier heights $V$, asymmetries $\Delta$, 
and distance $d$ between the wells, hence $\lambda = d \sqrt{m V/2\hbar^2}$) derives from the Gaussian distribution of $z$ and can be computed {\em 
analytically}. For larger $|z|$, the $U_{\rm RS}(v)$ only exhibit (anharmonic) 
single well forms; see Fig. 2. 

\begin{figure}[h]
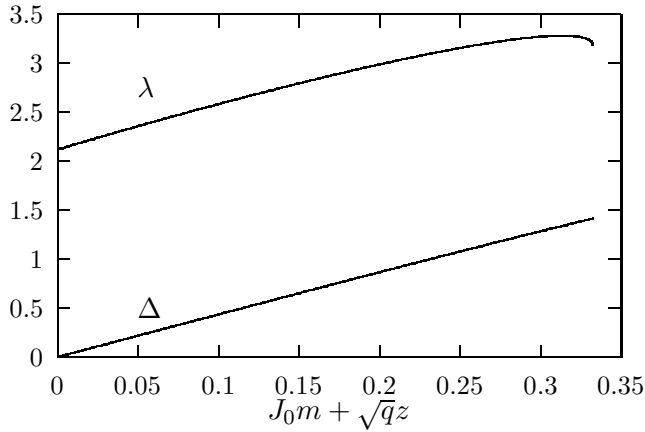

\setlength{\unitlength}{0.240900pt}
\ifx\plotpoint\undefined\newsavebox{\plotpoint}\fi
\sbox{\plotpoint}{\rule[-0.200pt]{0.400pt}{0.400pt}}%

\caption[]{Asymmetry $\Delta$ and $\lambda$ as functions of
 $\tilde h = J_0 m +\sqrt q z$. The parameters $\gamma$ and $J_0$ are as in
Fig. 2.}
\end{figure}

In contrast to the assumptions of the STM, we find that $\Delta$ and $\lambda$
are strongly correlated random variables; in the RS approximation both are
functions of one Gaussian random variable, viz. $z$, as shown in Fig. 3. The
distributions $P(\lambda)$ and $P(\Delta)$ are depicted in Fig. 4. Both have
singularities at their upper boundary, the former an integrable divergence, the
latter a cusp singularity. A notable feature here is that upper and lower limits
of the $\lambda$ and $\Delta$ ranges are {\em given\/} within our approach,
and the total mass under either distribution gives the fraction of degrees of
freedom which `see' DWPs.

\begin{figure}[h]
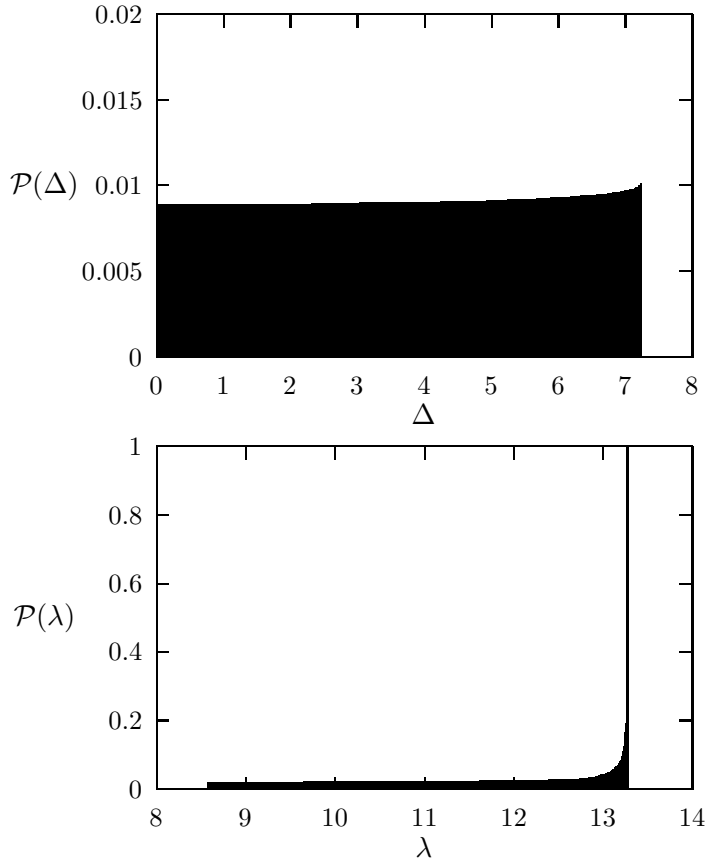

\setlength{\unitlength}{0.240900pt}
\ifx\plotpoint\undefined\newsavebox{\plotpoint}\fi
\sbox{\plotpoint}{\rule[-0.200pt]{0.400pt}{0.400pt}}%

\caption[]{Distributions ${\cal P}(\Delta)$ (for positive $\Delta$), and 
${\cal P}(\lambda)$. Here $\gamma = 4$, while $J_0$ is as in Fig. 2. The total 
mass under either distribution is 0.138 and gives the fraction of particles which `see' DWPs.}
\end{figure}

\begin{figure}[h]
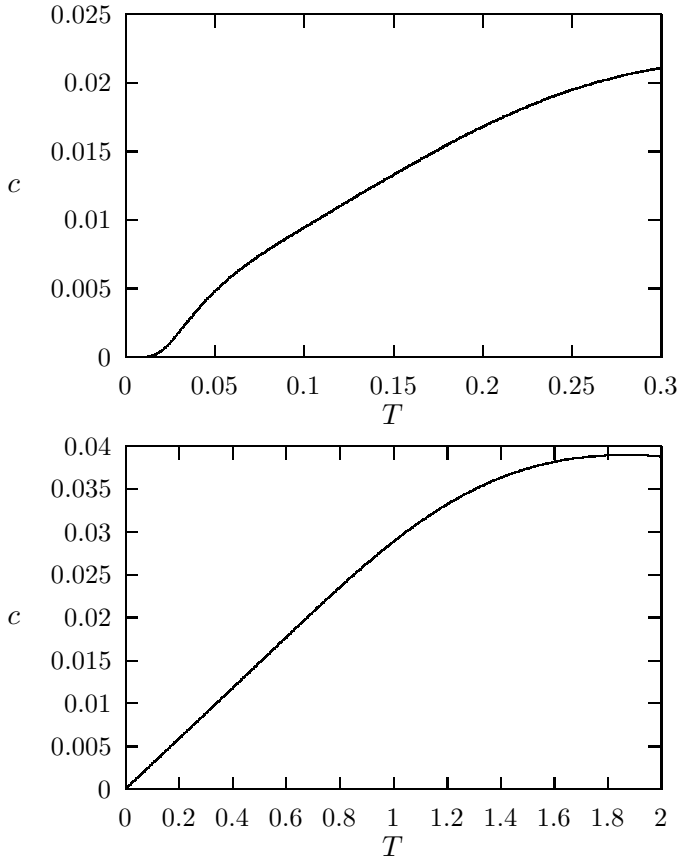

\setlength{\unitlength}{0.240900pt}
\ifx\plotpoint\undefined\newsavebox{\plotpoint}\fi
\sbox{\plotpoint}{\rule[-0.200pt]{0.400pt}{0.400pt}}%
%
\caption[]{Low $T$ specific heat. {\bf (a):} $\gamma = 2$, $J_0 = 0$. 
{\bf (b):} $\gamma = 4$, $J_0 = 0$. Note the extended linear region in {\bf 
(b)}. In both cases there is a region at very low $T$ where the behaviour is
exponential. In {\bf (b)} this region is too small to be resolved.}
\end{figure}

Despite the differences in the shape of $P(\Delta,\lambda)$ from that assumed
in the STM, we find that the contribution of the tunneling states to the
specific heat -- taking the tunnel splitting to be given by $\epsilon = 
\sqrt{\Delta^2 + \Delta_0^2}$ \cite{an+}, and ignoring higher excitations --
exhibits an extended range of temperatures where it scales linearly with $T$,
and we find this phenomenon to be more pronounced, as we move deeper into
the DWP phase, i.e., deeper also into the glassy phase as it is described in
our model (see Fig. 5). At very low temperatures exponential behaviour of the 
specific heat is observed which is due to the cutoff in the $\lambda$ 
distribution, which in turn creates a cutoff in the density of states at low energies.

\section{Summary and Outlook}

We have proposed and solved a simple model which exhibits both, a glass--like
freezing transition, and a collection of DWPs in its zero temperature potential
energy landscape. The latter are generally believed to give rise to a number
of low-temperature anomalies in glassy and amorphous systems via a tunneling
mechanism that allows particles to move back and forth between the wells forming
the DWP structure at temperatures where thermally activated classical motion
would still be rather unlikely. Within our model, we were able to compute the
distribution of the parameters characterizing the DWPs analytically, and we 
found, in particular, strong correlations between asymmetries $\Delta$ and the
parameter $\lambda$ which determines the magnitude of the tunneling matrix 
element $\Delta_0$. Nevertheless, we observe an extended range of low 
temperatures at which the contribution of the two-level tunneling systems to
the specific heat scales linearly with temperature. The correlations between
$\lambda$ and $\Delta$ can be weakened (but most likeley {\em not\/} eliminated)
by introducing {\em local\/} randomness, i.e., by introducing either a randomly 
$i$--dependent $\gamma$ or by making other parameters of the on--site potentials
$i$--dependent in a random fashion. It has been demonstrated elsewhere 
\cite{kub} that the model remains solvable with these modifications.

So far, we have evaluated only the replica symmetric approximation. At the same
time we know that RSB is observed in the interesting region of the phase diagram
where DWPs actually do occur. However, it can be shown that DWPs and even the
correlation between asymmetries and tunneling matrix element persist at all
finite levels of RSB \cite{hoku}; the distribution of the parameters characterizing them may of course change in details. The one- and two-step RSB approximations \cite{Pa80a} as well as Parisi's full RSB scheme \cite{Pa80b} for this model are currently being evaluated \cite{hoku}.

Concerning the motivation of (1) via the idea of an expansion of the potential
energy about a set of reference positions, it should be noted that in principle
a linear random--field type term should be added to $U_{\rm pot}$. While such a
term does change details of the low-temperature properties of the system, we 
find that the main physics is left invariant \cite{hoku}.

Up to now we have not investigated {\em relations\/} between high--temperature
and low--temperature properties of our system in any detail, but it should
be obvious that they exist --- all properties are determined from the two
model--parameters $\gamma$ and $J_0$ --- and that they are within relatively
easy reach of our approach; they are currently under study \cite{hoku}. 
Moreover, we have as yet treated the DWPs only as two--level systems, which 
entails that their contribution to the specific heat levels off and eventually 
decreases to zero as the temperature is further increased. Clearly, the 
contribution of higher excitations as well as that of the single well 
configurations to physical quantities must finally be taken into account as 
well.

It should be pointed out that our model will also  exhibit interesting
{\em dynamical\/} properties at or near its glass transition temperature and
very likely throughout the glassy phase, which are worth investigating. Indeed,
{\em formally\/} such an investigation already exists \cite{sozi}, albeit for
the Langevin dynamics of a model with $G$ chosen differently, so as to describe 
Ising spins. It would be interesting to see which features of that model will 
survive in the present context, and which will turn out to be altered. 

In the present paper we have chosen to quantize our system only {\em after\/}
a mean--field decoupling. For ordered, translationally invariant systems, the
validity of this procedure has been rigorously proven by Fannes et al.
\cite{fa+}. A corresponding proof for disordered systems is still lacking. 
Therefore, it would be interesting to study the system in a full quantum statistical context right from the outset, using imaginary time path integrals 
in conjunction with the replica method. It would  seem feasible to solve
the system along these lines at least within the so called static approximation.
The clear connection to the DWP concept, to which we have access precisely
through the mean--field decoupling scheme is, however, likely to be lost in a
solution along these lines.

\subsubsection* {Acknowledgements}

It is a pleasure to thank C. Enss, H. Horner, U. Horstmann, S. Hunklinger,
Th. Nieuwenhuizen, P. Neu, A. W\"urger, A. Verbeure, and in particular O. Terzidis for illuminating discussions and helpful advice.

\end{document}